\pdfoutput=1
\documentclass[12pt, epsfig]{article}
\usepackage{amsmath,amsfonts,amssymb,amsthm,amstext,amscd,eucal,graphicx,xcolor,endnotes}
\usepackage[all]{xy}
\usepackage{dsfont,epiolmec}
\usepackage{amsmath}
\usepackage{slashed,ccaption,cite}
\usepackage{mathrsfs, calligra}
\usepackage{ulem}
\usepackage[unicode=true,
 bookmarks=false,
 breaklinks=false,pdfborder={0 0 1},backref=false,colorlinks=true]
 {hyperref}
\hypersetup{
 pdfauthor={Daniel Grumiller, Shahin Sheikh-Jabbari },
 linkcolor=blue,urlcolor=magenta,filecolor=green,citecolor=red,pdfstartview=FitV,bookmarksopen=true}

\makeatletter \@addtoreset{equation}{section}

\makeatletter\renewcommand\section{\@startsection {section}{1}{\z@}%
                                   {-3.5ex \@plus -1ex \@minus -.2ex}%nn
                                   {2.3ex \@plus.2ex}%
                                   {\normalfont\large\bfseries}}
\renewcommand\subsection{\@startsection{subsection}{2}{\z@}%
                                     {-3.25ex\@plus -1ex \@minus -.2ex}%
                                     {1.5ex \@plus .2ex}%
                                     {\normalfont\bfseries}}

\renewcommand{\baselinestretch}{1.2}

\makeatletter
\DeclareFontFamily{OMX}{MnSymbolE}{}
\DeclareSymbolFont{MnLargeSymbols}{OMX}{MnSymbolE}{m}{n}
\SetSymbolFont{MnLargeSymbols}{bold}{OMX}{MnSymbolE}{b}{n}
\DeclareFontShape{OMX}{MnSymbolE}{m}{n}{
    <-6>  MnSymbolE5
   <6-7>  MnSymbolE6
   <7-8>  MnSymbolE7
   <8-9>  MnSymbolE8
   <9-10> MnSymbolE9
  <10-12> MnSymbolE10
  <12->   MnSymbolE12
}{}
\DeclareFontShape{OMX}{MnSymbolE}{b}{n}{
    <-6>  MnSymbolE-Bold5
   <6-7>  MnSymbolE-Bold6
   <7-8>  MnSymbolE-Bold7
   <8-9>  MnSymbolE-Bold8
   <9-10> MnSymbolE-Bold9
  <10-12> MnSymbolE-Bold10
  <12->   MnSymbolE-Bold12
}{}

\let\llangle\@undefined
\let\rrangle\@undefined
\DeclareMathDelimiter{\llangle}{\mathopen}%
                     {MnLargeSymbols}{'164}{MnLargeSymbols}{'164}
\DeclareMathDelimiter{\rrangle}{\mathclose}%
                     {MnLargeSymbols}{'171}{MnLargeSymbols}{'171}
\makeatletter
\newcommand{\be}{\begin{equation}}
\newcommand{\ee}{\end{equation}}
\newcommand{\bea}{\begin{eqnarray}}
\newcommand{\eea}{\end{eqnarray}}
\newcommand{\bse}{\begin{subequations}}
\newcommand{\ese}{\end{subequations}}
\newcommand{\beqa}{\begin{eqnarray}}
\newcommand{\eeqa}{\end{eqnarray}}
\newcommand{\beqar}{\begin{eqnarray*}}
\newcommand{\eeqar}{\end{eqnarray*}}
\newcommand{\bi}{\begin{itemize}}
\newcommand{\ei}{\end{itemize}}
\newcommand{\bn}{\begin{enumerate}}
\newcommand{\en}{\end{enumerate}}
\newcommand{\ba}{\begin{array}}
\newcommand{\ea}{\end{array}}
\newcommand{\bc}{\begin{center}}
\newcommand{\ec}{\end{center}}

\newcommand{\eq}[2]{\begin{equation} #1 \label{#2} \end{equation}}

\newcommand{\nhs}{${\boldsymbol{\mathcal{N}}}$}

\DeclareMathOperator{\extdm}{d}
\newcommand{\extd}{\extdm \!}

\definecolor{darkgreen}{rgb}{0,0.3,0}
\definecolor{darkblue}{rgb}{0,0,0.3}
\definecolor{darkred}{rgb}{0.7,0,0}

\newcommand{\old}[1]{}%{\sout{#1}}

\begin{document}
\renewcommand{\baselinestretch}{1.2}  %Line spacing

\newcommand\cnote[1]{\textcolor{red}{\bf [C:\,#1]}}
\newcommand\dnote[1]{\textcolor{magenta}{\bf [D:\,#1]}}
\newcommand\snote[1]{\textcolor{blue}{\bf [S:\,#1]}}

\begin{titlepage}

\begin{flushright}%\vspace{-3cm}
{
TUW--20--20\\ %, IPM/P-20/xxx\\
March 31, 2020 }\end{flushright}

\vspace*{2truecm}

\newcommand{\mytitle}{Horizons 2020}

\begin{center}
\LARGE{\bf{\mytitle}}

\vspace*{2truecm}

\large{\bf{D.~Grumiller\footnote{Corresponding author; e-mail:~\href{grumil@hep.itp.tuwien.ac.at}{grumil@hep.itp.tuwien.ac.at}}$^{; a}$, M.M.~Sheikh-Jabbari\footnote{e-mail:~\href{jabbari@theory.ipm.ac.ir}{jabbari@theory.ipm.ac.ir}}$^{; b,c}$} and C.~Zwikel\footnote{e-mail:~\href{zwikel@hep.itp.tuwien.ac.at}{zwikel@hep.itp.tuwien.ac.at}}$^{; a}$}
\\

\normalsize
\bigskip

{$^a$ \it Institute for Theoretical Physics, TU Wien\\
Wiedner Hauptstr.~8, A-1040 Vienna, Austria}
\\
{$^b$ \it School of Physics, Institute for Research in Fundamental
Sciences (IPM)\\ P.O.Box 19395-5531, Tehran, Iran}\\
{$^c$ \it The Abdus Salam ICTP, Strada Costiera 11, Trieste, Italy}

\end{center}
\setcounter{footnote}{0}

\bigskip

% ABSTRACT WORD LIMIT: 125 WORDS; currently: 99 words

\begin{abstract}
Horizons of black holes or cosmologies are peculiar loci of spacetime where interesting physical effects takes place, some of which are probed by recent (EHT and LIGO) and future experiments (ET and LISA). We discuss that there are boundary degrees of freedom residing at the horizon. We describe their symmetries and their interactions with gravitational waves. This fits into a larger picture of boundary plus bulk degrees of freedom and their interactions in gauge theories. Existence and dynamics of the near horizon degrees of freedom could be crucial to address fundamental questions and apparent paradoxes in black holes physics.
\end{abstract}

\vfill

\centerline{\textit{Essay written for the Gravity Research Foundation 2020 Awards for Essays on Gravitation.}}

\end{titlepage}

\newpage

%%%%%%%%%%%%%%%%%%%%%%%%%%%%%%%%%%%%%%%%%%%%%%%%%%%%%%%%%%%%%%%%%%%%%%%%%%%%%%%%
%%%                                                                          %%%
%%% MAXIMALLY THREE PAGES OF TEXT WITH CURRENT LAYOUT (LESS THAN 1500 WORDS) %%%
%%%                                                                          %%%
%%%%%%%%%%%%%%%%%%%%%%%%%%%%%%%%%%%%%%%%%%%%%%%%%%%%%%%%%%%%%%%%%%%%%%%%%%%%%%%%

%\vspace*{0.1truecm}

% MAIN TEXT WORD LIMIT: 1500 WORDS; current word count: about 1600, but including formulas and citations; so probably we are just fine

% QUOTE ``Infinite horizons belong to those who have infinite imagination!'', Mehmet Murat ildan 
% QUOTE ``No one should ever be deprived of a horizon.'', Maude Julien
% QUOTE ``Build your own horizons!'', Suyasha Subedi 

%{\LARGE{\calligra{H}}}
Horizons have undergone a fascinating history of conceptual changes, some of them going back-and-forth. The Schwarzschild black hole horizon initially was considered as singular. By 1960 it was evident that such singularities are coordinate artifacts. Until the 1970ies horizons were not considered to have physical properties (`no hair') and were merely thought of as (regular) boundaries of a causal patch, e.g.~for an observer outside a black hole, {as implied by a strict reading of Einstein's equivalence principle}. This perception gradually changed in the last quarter of the past century thanks to Hawking's discovery that black hole horizons come with a temperature and entropy, and eventually culminated in the membrane paradigm that attached further physical properties, such as viscosity or electrical conductivity, to black hole horizons \cite{Thorne:1986iy}. 

While the classical interpretation of horizons became clearer and physically richer in that period, it was realized that their quantum interpretation is puzzling \cite{Hawking:1976ra,tHooft:1984kcu}. Especially the microscopic derivation of the Bekenstein--Hawking entropy associated with black hole horizons and the information paradox became attractors for numerous research avenues, ranging from ambitious to quixotic. Attempts to resolve these puzzles often involve developing ideas around horizons: they became stretched \cite{tHooft:1990fkf,Susskind:1993if}, isolated \cite{Ashtekar:2000sz}, equipped with brick-walls \cite{'tHooft:2004ek} or soft hair \cite{Hawking:2016msc}, entangled with the interior \cite{Page:1993wv,Hayden:2007cs}, converted into fuzzballs \cite{Mathur:2005zp} or fluffballs \cite{Afshar:2017okz}, identified antipodally \cite{Hooft:2016vug} and, like an echo from the past, were reconsidered as singular surfaces due to the appearance of firewalls \cite{Almheiri:2012rt}. If firewalls did exist a cornerstone of classical gravity, the equivalence principle, could no longer be true.

Our 2020 viewpoint on horizons is not quite as radical, but rather a blend of some of the proposals mentioned above \cite{Grumiller:2019fmp,Adami:2020amw}. The equivalence principle in presence of horizons (or general boundaries) indeed needs some refinement \cite{Sheikh-Jabbari:2016lzm}, since there are diffeomorphisms, often referred to as `improper diffeomorphisms', that have physical effects. By contrast, the proper ones account for gauge redundancies. This is not particular to gravity or horizons, but applies to any gauge field theory with an actual, fiducial or asymptotic boundary. 

As a consequence of improper diffeomorphisms there are degrees of freedom (d.o.f.) residing at the boundary. In principle the microscopic details of the boundary specify both the boundary d.o.f.\ and their interaction with the environment. Alternatively, in a macroscopic effective description, the boundary d.o.f.\ and their interactions can be determined through appropriate boundary conditions on the bulk fields. Archetypical examples of such boundary conditions in electrodynamics are the Casimir effect or the description of conductors as equipotential surfaces, see e.g.\ \cite{Barnich:2019xhd,Barnich:2019qex}. That asymptotic flat geometries are specified up to the Bondi news \cite{Bondi:1962, Sachs:1962} is a famous gravity example of physical effects associated with (asymptotic) boundary conditions.

Since our interest is in horizons we count the d.o.f.\ using a $(d-2)+2$ split, using the labels $x^\pm$ for the null directions and $x^A$ for the transversal directions. As a warm-up consider electrodynamics, where the gauge connection splits into a vector $A_A$ and two scalars $A_+,\,A_-$. Gauge-fixing $A_-=0$ leads to $d-1$ d.o.f., $d-2$ of which are the usual massless photon polarizations, which we refer to as bulk d.o.f. The remaining d.o.f.\ $A_+$ can be changed arbitrarily by residual gauge transformations that preserve the gauge choice $A_-=0$, namely $A_+\to A_+ + \partial_+\lambda(x^+,\,x^A)$, and hence has no local influence on physics. However, it is a physical boundary d.o.f. Thus, in electrodynamics we have $d-2$ bulk d.o.f. and one boundary d.o.f., which is an arbitrary function of $x^+,x^A$. In certain applications boundary conditions are imposed that partially eliminate  bulk or boundary d.o.f., so when we make statements like `electrodynamics has one boundary d.o.f.' we assume that there is not such a restriction of the configuration space. 

In general relativity we split the metric into a traceless co\-di\-men\-sion-two tensor $g_{AB}^T$, two co\-di\-men\-sion-two vectors $g_{A\pm}$ and four scalars, $g_{\pm\pm},\,g_{+-}$ and $g^T$, where the latter is the trace-part of $g_{AB}$. We use the $d$ diffeomorphisms to gauge fix one of the vectors and two of the scalars. The d.o.f.\ counting then works as follows. The bulk d.o.f.\ are contained in the traceless tensor, which has $(d-1)(d-2)/2-1=d(d-3)/2$ independent components, namely the polarizations of massless gravitons. The remaining d.o.f., $d(d+1)/2-d(d-3)/2-d=d$, can be altered by residual diffeomorphisms which preserve our gauge choice, but again are physical at the boundary. Thus, in general relativity we have $d(d-3)/2$ bulk d.o.f.\ and $d$ boundary d.o.f. A simple example is three-dimensional gravity which has zero bulk d.o.f. Its three boundary d.o.f.\ were identified in \cite{Grumiller:2016pqb} and the seminal Brown--Henneaux construction  \cite{Brown:1986nw} only realizes one of them. In our essay we focus on the physically most interesting case $d=4$, with particular emphasis on horizons. 

Consider a null hypersurface \nhs\ located at $x^-=0$ so that $x^-$ is a small expansion parameter (we demand Taylor expandability of the metric in $x^-$). Having a null hypersurface means $g^{--}={\cal O}(x^-)$. We choose $\partial_-$ to be hypersurface orthogonal and the other null direction to be along $\partial_+$ at $x^-=0$. %$x^+$ may be identified as lightlike `time' along \nhs. 
Our configuration space consists of all metrics compatible with the near null hypersurface behavior
\eq{
\extd s^2 =  \eta\,\extd x^+\extd x^- + \Omega_{AB}\,\extd x^A \extd x^B + x^-\,\big({\mathcal G} \,\extd x^{+\,2} + \theta_A \,\extd x^+\extd x^A + \lambda_{AB}\,\extd x^A \extd x^B\big) + {\cal O}(x^{-\,2})
}{eq:1}
where we partially gauge-fixed $g_{--}=0=g_{-A}$, which turns out to be possible with no loss of generality \cite{Adami:2020amw}. All functions in the metric \eqref{eq:1} can depend on `time' $x^+$ along \nhs\ and `angular coordinates' $x^A$ transversal to \nhs. 

In accordance with our general discussion the information about bulk d.o.f.\ resides in the traceless parts of $\Omega_{AB}$ and $\lambda_{AB}$, which describe gravitational waves (GWs). The boundary d.o.f.\ are described by the codimension-two vector $\theta_A$ and two scalars, the boost function $\eta$ and the expansion function $\Omega :=\sqrt{\det{\Omega_{AB}}}$. The additional scalar $\mathcal G$ and the trace part of $\lambda_{AB}$ present in \eqref{eq:1} must be redundant; {the Einstein equations indeed determine them in terms of the other functions in \eqref{eq:1}.}
In summary, we have two bulk d.o.f.\ in the form of GWs and four boundary d.o.f. Thus, our near null hypersurface expansion \eqref{eq:1} is general enough to accommodate essentially the maximum of boundary d.o.f.\footnote{%
We included the qualifier `essentially' because $\partial_+\theta_A$ is determined in terms of $\eta, \Omega_{AB}$ through Einstein's equations. So only the $x^+$ independent part of $\theta_A$ is among our boundary d.o.f.\ \cite{Adami:2020amw}. This could be relaxed by dropping the assumption of hypersurface orthogonality.
} Bulk and boundary d.o.f.\ are dynamically coupled to each other through the equations of motion, leading to a rich phenomenology of black holes and their boundary excitations interacting with bulk GWs.

The near null hypersurface expansion \eqref{eq:1} is preserved by near null hypersurface Killing vectors 
\eq{
\xi[T^\pm, Y^A]=T^+(x^+,x^B)\,\partial_+ - x^- T^-(x^+,x^B)\,\partial_- + Y^A(x^B)\,\partial_A +{\cal O}((x^-)^2)\,,
}{eq:nhkv}
where $T^\pm(x^+,x^A),\,Y^A(x^B)$ are arbitrary functions of their arguments. By `preserved' we mean that they map any metric of the form \eqref{eq:1} to another metric of the same form. The vectors \eqref{eq:nhkv} thus are a pendant to the asymptotic Killing vectors of Bondi, van der Burgh, Metzner and Sachs \cite{Bondi:1962, Sachs:1962}. 

The near null hypersurface Killing vectors with the usual Lie bracket close into an algebra.  They also lead to boundary (surface) charges $Q_\xi$ which are given by integrals over the codimension-two $x^A$ part and are functionals of boundary d.o.f. These charges  in general are not conserved since we can have fluxes $F$ through the horizon. They are related by a generalized charge conservation equation
\begin{equation}
{\partial_+}Q_\xi=-F
%-{ F}_{\partial_+}(\delta_\xi g) 
\label{GCCE}
\end{equation}
reminiscent of the continuity equation in electrodynamics. The explicit form of the charges, their algebra and the fluxes can be found in \cite{Adami:2020amw}. 
 
Depending on the physical question, one might need to restrict the configuration space by imposing boundary conditions. %\footnote{\tcb{In the `asymptotic symmetry analysis', these boundary conditions are typically imposed through certain fall-off conditions. As our discussions makes it clear, boundary conditions is more general than specifying fall-off behavior}.} 
This in turn can switch off or restrict certain boundary d.o.f. For example, allowing only linear $x^+$ dependence in the expansion function $\Omega$ fixes the boost function $\eta=2$. The latter choice has been the starting point of any analysis that adopts Gaussian null coordinates, which is rather common in the literature on null hypersurfaces, see \cite{Donnay:2015abr,Donnay:2016ejv} for such a choice in a near horizon context. An important lesson from this example is that gauge fixing to Gaussian null coordinates is with loss of generality, as it switches off one of the boundary d.o.f. Similarly, the well-known Fefferman--Graham expansion in asymptotically anti-de~Sitter switches off some boundary d.o.f., and thus is also with loss of generality \cite{Grumiller:2016pqb}. Therefore, if one intends to keep the maximal number of boundary d.o.f.\ one should avoid premature or inadequate gauge-fixings that reduce this number. 

Let us address a final issue, namely field redefinitions. In our example \eqref{eq:1} we may choose to describe the configuration space through new configuration variables $\tilde\eta,\,\tilde\Omega,\,\tilde\theta_A$, which are generic functions of the original ones, $\eta,\,\Omega,\,\theta_A$. In other words, one may slice the configuration space in different ways. This is conceptually similar to Legendre-transformation between different thermodynamical ensembles and will have an impact on the field dependence of the symmetry generators, as they may be viewed as generators of motion in the configuration space with respect to a given slicing. Thus, such field redefinitions can modify the symmetry algebra. A near horizon example (in the truncated stationary setup \cite{Adami:2020amw}) is the redefinition $\tilde\Omega=\Omega^{r+1}/(r+1)$, where $r=s/(d-2)$ with some non-negative integer $s$. As discussed in \cite{Grumiller:2019fmp}, this redefinition leads to near horizon supertranslations of spin $s$ and for $s=1$ provides an explicit near horizon realization of BMS symmetries. 

In conclusion, our understanding of horizons has increased over the past few years by considering near horizon expansions that allow an increasing number of physical boundary d.o.f.\ The maximal number of four boundary d.o.f.\ was only reached in 2020.

\bigskip

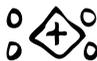
\begin{center}
{\LARGE{\EOstar}}
\end{center}

%%%%%%%%%%%%%%%%%%%%%%%%%%%%%%%%%%%%%%%%%%%%%%%%%%%%%%%%%
%%%                                                   %%%
%%% WORD COUNT OF 1500 IS SATURATED IF WE USE 3 PAGES %%%
%%%                                                   %%%
%%%%%%%%%%%%%%%%%%%%%%%%%%%%%%%%%%%%%%%%%%%%%%%%%%%%%%%%%

\section*{Acknowledgments}

We thank Hamed Adami, Alfredo P\'erez, Saeedeh Sadeghian, C\'edric Troessaert, Ricardo Troncoso and Raphaela Wutte for recent collaboration on near horizon symmetries and Laura Donnay for discussions.
DG was supported by the Austrian Science Fund (FWF), projects P~30822 and P~32581. 
MMShJ acknowledge the support by INSF grant No.~950124 and Saramadan grant No.~ISEF/M/98204. 
CZ was supported by the Austrian Science Fund (FWF), projects P~30822 and M~2665. 
We acknowledge the Iran-Austria IMPULSE project grant, supported and run by Khawrizmi University and OeAD.

% make refs. appear in italics, not underline
\renewcommand{\em}{\it}

\bibliographystyle{fullsort}
%\bibliography{reference}

\begin{thebibliography}{10}

\bibitem{Thorne:1986iy}
K.~S. Thorne, R.~Price, and D.~Macdonald, {\em Black Holes: The Membrane
  Paradigm}.
\newblock Yale University Press,
1986.
\newblock
%%CITATION = ISBN-9780300037708 ETC.;%%.

\bibitem{Hawking:1976ra}
S.~W. Hawking, ``Breakdown of predictability in gravitational collapse,'' {\em
  Phys. Rev.} {\bf D14} (1976)
2460--2473.
%%CITATION = PHRVA,D14,2460;%%.

\bibitem{tHooft:1984kcu}
G.~'t~Hooft, ``{On the Quantum Structure of a Black Hole},'' {\em Nucl. Phys.}
  {\bf B256} (1985)
727--745.
%%CITATION = NUPHA,B256,727;%%.

\bibitem{tHooft:1990fkf}
G.~'t~Hooft, ``{The black hole interpretation of string theory},'' {\em Nucl.
  Phys.} {\bf B335} (1990)
138--154.
%%CITATION = NUPHA,B335,138;%%.

\bibitem{Susskind:1993if}
L.~Susskind, L.~Thorlacius, and J.~Uglum, ``{The Stretched horizon and black
  hole complementarity},'' {\em Phys.Rev.} {\bf D48} (1993) 3743--3761,
\href{http://www.arXiv.org/abs/hep-th/9306069}{{\tt hep-th/9306069}}.
%%CITATION = HEP-TH/9306069;%%.

\bibitem{Ashtekar:2000sz}
A.~Ashtekar {\em et al.}, ``Isolated horizons and their applications,'' {\em
  Phys. Rev. Lett.} {\bf 85} (2000) 3564--3567,
\href{http://www.arXiv.org/abs/gr-qc/0006006}{{\tt gr-qc/0006006}}.
%%CITATION = GR-QC 0006006;%%.

\bibitem{'tHooft:2004ek}
G.~'t~Hooft, ``Horizons,''
\href{http://www.arXiv.org/abs/gr-qc/0401027}{{\tt gr-qc/0401027}}.
%%CITATION = GR-QC 0401027;%%.

\bibitem{Hawking:2016msc}
S.~W. Hawking, M.~J. Perry, and A.~Strominger, ``{Soft Hair on Black Holes},''
  {\em Phys. Rev. Lett.} {\bf 116} (2016), no.~23, 231301,
\href{http://www.arXiv.org/abs/1601.00921}{{\tt 1601.00921}}.
%%CITATION = ARXIV:1601.00921;%%.

\bibitem{Page:1993wv}
D.~N. Page, ``{Information in black hole radiation},'' {\em Phys. Rev. Lett.}
  {\bf 71} (1993) 3743--3746,
\href{http://www.arXiv.org/abs/hep-th/9306083}{{\tt hep-th/9306083}}.
%%CITATION = HEP-TH/9306083;%%.

\bibitem{Hayden:2007cs}
P.~Hayden and J.~Preskill, ``{Black holes as mirrors: Quantum information in
  random subsystems},'' {\em JHEP} {\bf 09} (2007) 120,
\href{http://www.arXiv.org/abs/0708.4025}{{\tt 0708.4025}}.
%%CITATION = ARXIV:0708.4025;%%.

\bibitem{Mathur:2005zp}
S.~D. Mathur, ``{The Fuzzball proposal for black holes: An Elementary
  review},'' {\em Fortsch.Phys.} {\bf 53} (2005) 793--827,
\href{http://www.arXiv.org/abs/hep-th/0502050}{{\tt hep-th/0502050}}.
%%CITATION = HEP-TH/0502050;%%.

\bibitem{Afshar:2017okz}
H.~Afshar, D.~Grumiller, M.~M. Sheikh-Jabbari, and H.~Yavartanoo, ``{Horizon
  fluff, semi-classical black hole microstates --- Log-corrections to BTZ
  entropy and black hole/particle correspondence},'' {\em JHEP} {\bf 08} (2017)
  087,
\href{http://www.arXiv.org/abs/1705.06257}{{\tt 1705.06257}}.
%%CITATION = ARXIV:1705.06257;%%.

\bibitem{Hooft:2016vug}
G.~'t~Hooft, ``{The Firewall Transformation for Black Holes and Some of Its
  Implications},'' {\em Found. Phys.} {\bf 47} (2017), no.~12, 1503--1542,
\href{http://www.arXiv.org/abs/1612.08640}{{\tt 1612.08640}}.
%%CITATION = ARXIV:1612.08640;%%.

\bibitem{Almheiri:2012rt}
A.~Almheiri, D.~Marolf, J.~Polchinski, and J.~Sully, ``{Black Holes:
  Complementarity or Firewalls?},'' {\em JHEP} {\bf 1302} (2013) 062,
\href{http://www.arXiv.org/abs/1207.3123}{{\tt 1207.3123}}.
%%CITATION = ARXIV:1207.3123;%%.

\bibitem{Grumiller:2019fmp}
D.~Grumiller, A.~P{\'e}rez, M.~M. Sheikh-Jabbari, R.~Troncoso, and C.~Zwikel,
  ``{Spacetime structure near generic horizons and soft hair},'' {\em Phys.
  Rev. Lett.} {\bf 124} (2020) 041601,
\href{http://www.arXiv.org/abs/1908.09833}{{\tt 1908.09833}}.
%%CITATION = ARXIV:1908.09833;%%.

\bibitem{Adami:2020amw}
H.~Adami, D.~Grumiller, S.~Sadeghian, M.~M. Sheikh-Jabbari, and C.~Zwikel,
  ``{T-Witts from the horizon},''
\href{http://www.arXiv.org/abs/2002.08346}{{\tt 2002.08346}}.
%%CITATION = ARXIV:2002.08346;%%.

\bibitem{Sheikh-Jabbari:2016lzm}
M.~M. Sheikh-Jabbari, ``{Residual diffeomorphisms and symplectic soft hairs:
  The need to refine strict statement of equivalence principle},'' {\em Int. J.
  Mod. Phys.} {\bf D25} (2016), no.~12, 1644019,
\href{http://www.arXiv.org/abs/1603.07862}{{\tt 1603.07862}}.
%%CITATION = ARXIV:1603.07862;%%.

\bibitem{Barnich:2019xhd}
G.~Barnich, ``{Black hole entropy from nonproper gauge degrees of freedom: The
  charged vacuum capacitor},'' {\em Phys. Rev.} {\bf D99} (2019), no.~2,
  026007,
\href{http://www.arXiv.org/abs/1806.00549}{{\tt 1806.00549}}.
%%CITATION = ARXIV:1806.00549;%%.

\bibitem{Barnich:2019qex}
G.~Barnich and M.~Bonte, ``{Soft degrees of freedom, Gibbons-Hawking
  contribution and entropy from Casimir effect},''
\href{http://www.arXiv.org/abs/1912.12698}{{\tt 1912.12698}}.
%%CITATION = ARXIV:1912.12698;%%.

\bibitem{Bondi:1962}
H.~Bondi, M.~van~der Burg, and A.~Metzner, ``Gravitational waves in general
  relativity {VII.} {W}aves from axi-symmetric isolated systems,'' {\em Proc.
  Roy. Soc. London} {\bf A269} (1962) 21--51.

\bibitem{Sachs:1962}
R.~Sachs, ``Asymptotic symmetries in gravitational theory,'' {\em Phys. Rev.}
  {\bf 128} (1962) 2851--2864.

\bibitem{Grumiller:2016pqb}
D.~Grumiller and M.~Riegler, ``{Most general AdS$_{3}$ boundary conditions},''
  {\em JHEP} {\bf 10} (2016) 023,
\href{http://www.arXiv.org/abs/1608.01308}{{\tt 1608.01308}}.
%%CITATION = ARXIV:1608.01308;%%.

\bibitem{Brown:1986nw}
J.~D. Brown and M.~Henneaux, ``{Central Charges in the Canonical Realization of
  Asymptotic Symmetries: An Example from Three-Dimensional Gravity},'' {\em
  Commun. Math. Phys.} {\bf 104} (1986)
207--226.
%%CITATION = CMPHA,104,207;%%.

\bibitem{Donnay:2015abr}
L.~Donnay, G.~Giribet, H.~A. Gonz{\'a}lez, and M.~Pino, ``{Supertranslations
  and Superrotations at the Black Hole Horizon},'' {\em Phys. Rev. Lett.} {\bf
  116} (2016), no.~9, 091101,
\href{http://www.arXiv.org/abs/1511.08687}{{\tt 1511.08687}}.
%%CITATION = ARXIV:1511.08687;%%.

\bibitem{Donnay:2016ejv}
L.~Donnay, G.~Giribet, H.~A. Gonz{\'a}lez, and M.~Pino, ``{Extended Symmetries
  at the Black Hole Horizon},'' {\em JHEP} {\bf 09} (2016) 100,
\href{http://www.arXiv.org/abs/1607.05703}{{\tt 1607.05703}}.
%%CITATION = ARXIV:1607.05703;%%.

\end{thebibliography}

\providecommand{\href}[2]{#2}\begingroup\raggedright\endgroup

\newpage
\end{document}